\documentclass[prd,aps,singlecolumn,preprintnumbers, showpacs, nofootinbib,superscriptaddress]{revtex4-1}

 \usepackage[T1]{fontenc}
 \usepackage[utf8]{inputenc}
 \usepackage{enumitem}
 \usepackage{color} 
 \usepackage{fancyhdr}
 \usepackage{physics}
 \usepackage{caption}
 \usepackage{amsmath}
 \usepackage{slashed}
 \usepackage{float}
 \usepackage{graphicx} 
 \usepackage{subcaption} 
 
\newcommand{\Blue}[1]{\textcolor{black}{#1}}

 \graphicspath{ {./images/} }

\begin{document}

\preprint{}

\title{
Chiral Perturbation for Large Momentum Effective Field Theory
}
\author{Wei-Yang Liu}
\email{r07222052@g.ntu.edu.tw}
\affiliation{Department of Physics, Center for Theoretical Physics, and Leung Center for Cosmology and Particle Astrophysics, National Taiwan University, Taipei, Taiwan 106}

\author{Jiunn-Wei Chen}
\email{jwc@phys.ntu.edu.tw}
\affiliation{Department of Physics, Center for Theoretical Physics, and Leung Center for Cosmology and Particle Astrophysics, National Taiwan University, Taipei, Taiwan 106}
\affiliation{Physics Division, National Center for Theoretical Sciences, National Taiwan University, Taipei 10617, Taiwan}

 \begin{abstract}
Large momentum effective field theory (LaMET) enables the extraction of parton distribution functions (PDF's) directly on a Euclidean lattice through a factorization theorem that relates the computed quasi-PDF's to PDF's. We apply chiral perturbation theory (ChPT) to LaMET to further separate soft scales, such as light quark masses and lattice size, to obtain leading model independent extrapolation formulas for extrapolations to physical quark masses and infinite volume. We find that the finite volume effect is reduced when the nucleon carries a finite momentum.
For nucleon momentum greater than 1 GeV and the lattice size $L$ and pion mass $m_{\pi}$ satisfying $m_{\pi} L \ge 3$, the finite volume effect is less than $1\%$ and is negligible for the current precision of lattice computations. This can be interpreted as a Lorentz contraction of the nucleon size
 in the z-direction which makes the lattice size effectively larger in that direction. We also find that the quark mass dependence in the infinite volume limit computed with non-zero nucleon momentum reproduces the previous result computed at zero momentum, as expected. Our approach can be generalized to other parton observables in LaMET straight forwardly.  
 \end{abstract}
 \maketitle
 
 \section{Introduction}
 
Large momentum effective theory (LaMET) enables computations of parton distribution functions (PDF's)  of hadrons on a Euclidean lattice. LaMET relates equal-time spatial correlators (whose Fourier transforms are called quasi-PDF's) to PDF's in the infinite hadron momentum limit~\cite{Ji:2013dva,Ji:2014gla}. For large but finite momenta accessible on a realistic lattice, LaMET relates quasi-PDF's to physical ones through a factorization theorem, which involves a matching coefficient and power corrections that are suppressed by the hadron momentum: 
\begin{equation}\label{eq:matching}
\tilde{q}(x, P_z, a)=\int_{-1}^1 \frac{dy}{|y|} Z(\frac{x}{y},y P_z,a,\mu) q(y, \mu) +\mathcal{O}(\frac{M^{2}}{P_z^2},\frac{\Lambda_{QCD}^{2}}{P_z^2}),
\end{equation}
where $x$ and $y$ are momentum fractions of the parton with respect to the hadron, $P_{z}$ is the hadron momentum, $\tilde{q}$ is the quasi-PDF defined computed on a lattice with lattice spacing $a$ while $\tilde{q}$ is the PDF defined at the scale $\mu$. $M$ is the hadron mass and  $\Lambda_{\text{QCD}}$ is the strong interaction scale. The hierarchy of scales follows
\begin{equation}
\frac{\pi}{a} \gg P_{z} \gg M, \Lambda_{QCD} \gg \frac{\pi}{L} ,
\end{equation}
with $L$ is the size of the lattice, such that the power correction is small.  $\tilde{q}$ and $q$ have the same infrared physics. Their difference in the ultraviolet is compensated by the matching kernel $Z$. The proof of factorization was developed in  Refs.~\cite{Ma:2017pxb,Izubuchi:2018srq,Liu:2019urm}.

Since LaMET was proposed, a lot of progress has been made in the theoretical understanding of the formalism~\cite{Xiong:2013bka,Ji:2015jwa,Ji:2015qla,Xiong:2015nua,Ji:2017rah,Monahan:2017hpu,Stewart:2017tvs,Constantinou:2017sej,Green:2017xeu,Izubuchi:2018srq,Xiong:2017jtn,Wang:2017qyg,Wang:2017eel,Xu:2018mpf,Chen:2016utp,Zhang:2017bzy,Ishikawa:2016znu,Chen:2016fxx,Ji:2017oey,Ishikawa:2017faj,Chen:2017mzz,Alexandrou:2017huk,Constantinou:2017sej,Green:2017xeu,Chen:2017mzz,Chen:2017mie,Lin:2017ani,Chen:2017lnm,Li:2016amo,Monahan:2016bvm,Radyushkin:2016hsy,Rossi:2017muf,Carlson:2017gpk,Ji:2017rah,Briceno:2018lfj,Hobbs:2017xtq,Jia:2017uul,Xu:2018eii,Jia:2018qee,Spanoudes:2018zya,Rossi:2018zkn,Liu:2018uuj,Ji:2018waw,Bhattacharya:2018zxi,Radyushkin:2018nbf,Zhang:2018diq,Li:2018tpe,Braun:2018brg,Detmold:2019ghl,Sufian:2020vzb,Shugert:2020tgq,Green:2020xco,Braun:2020ymy,Lin:2020ijm,Bhat:2020ktg,Chen:2020arf,Ji:2020baz,Chen:2020iqi,Chen:2020ody,Alexandrou:2020tqq,Fan:2020nzz,Ji:2020brr}. The method has been applied in lattice calculations of PDF's for the nucleon~\cite{Lin:2014zya,Chen:2016utp,Lin:2017ani,Alexandrou:2015rja,Alexandrou:2016jqi,Alexandrou:2017huk,Chen:2017mzz,Lin:2018pvv,Alexandrou:2018pbm,Chen:2018xof,Alexandrou:2018eet,Lin:2018qky,Fan:2018dxu,Liu:2018hxv,Wang:2019tgg,Lin:2019ocg,Liu:2020okp,Lin:2019ocg,Zhang:2019qiq,Alexandrou:2020qtt}, $\pi$~\cite{Chen:2018fwa,Izubuchi:2019lyk,Gao:2020ito} and $K$~\cite{Lin:2020ssv}  mesons. Despite limited volumes and relatively coarse lattice spacings, the state-of-the-art nucleon isovector quark PDF's, determined from lattice data at the physical point, have shown reasonable agreement~\cite{Lin:2018pvv,Alexandrou:2018pbm} with phenomenological results extracted from the experimental data. Encouraged by this success, LaMET has also been applied to $\Delta^{+}$~\cite{Chai:2020nxw} and twist-three PDF's~\cite{Bhattacharya:2020cen,Bhattacharya:2020xlt,Bhattacharya:2020jfj}, as well as gluon \cite{Fan:2020cpa}, strange and charm distributions~\cite{Zhang:2020dkn}. It was also applied to meson distribution amplitudes~\cite{Zhang:2017bzy,Chen:2017gck,Zhang:2020gaj,Hua:2020gnw} and
generalized parton distributions (GPD's)~\cite{Chen:2019lcm,Alexandrou:2020zbe,Lin:2020rxa,Alexandrou:2019lfo}. More recently, attempts have been made to generalize LaMET to transverse momentum dependent (TMD) PDF's~\cite{Ji:2014hxa,Ji:2018hvs,Ebert:2018gzl,Ebert:2019okf,Ebert:2019tvc,Ji:2019sxk,Ji:2019ewn,Ebert:2020gxr} to calculate the nonperturbative Collins-Soper evolution kernel~\cite{Ebert:2018gzl,Shanahan:2019zcq,Shanahan:2020zxr} and soft functions \cite{Zhang:2020dbb} on the lattice.
LaMET also brought renewed interests in earlier approaches~\cite{Liu:1993cv,Detmold:2005gg,Braun:2007wv,Bali:2017gfr,Bali:2018spj,Detmold:2018kwu,Liang:2019frk} and inspired new ones~\cite{Ma:2014jla,Ma:2014jga,Chambers:2017dov,Radyushkin:2017cyf,Orginos:2017kos,Radyushkin:2017lvu,Radyushkin:2018cvn,Zhang:2018ggy,Karpie:2018zaz,Joo:2019jct,Radyushkin:2019owq,Joo:2019bzr,Balitsky:2019krf,Radyushkin:2019mye,Joo:2020spy,Can:2020sxc}. For recent reviews, see, e.g., Refs. \cite{Lin:2017snn,Cichy:2018mum,Zhao:2020vll,Ji:2020ect,Ji:2020byp}. The renormalon ambiguity in the matching kernel $Z$ was first studied in Ref.~\cite{Braun:2018brg} which implies the power correction due to higher twist effect should be $\mathcal{O}(\Lambda^2_{\text{QCD}}/x^{2}P_z^2)$ to cancel the renormalon ambiguity. However, the study of bubble chain diagrams of Ref.~\cite{Liu:2020rqi} did not find the slow convergence of the kernel at three loop order, indicating that the renormalon effect could be mild to this order in quasi-PDFs. 

Despite the progress made in LaMET, the LaMET factorization theorem of Eq. (\ref{eq:matching}) makes no attempt to separate the light quark mass scales $m_{{u,d}}$ and $L$ from scales such as $\Lambda_{QCD}$ or $\Lambda_{\chi}$ (the scale of chiral symmetry breaking). Thus $\tilde{q}$ is a function of all these scales, while $q$ is expected to have the same quark mass dependence as $\tilde{q}$ but no volume dependence. As lattice exploration of the $m_{{u,d}}$ and $L$ parameter space requires a significant amount of computing resources, model independent formulas to guide the extrapolations to physical quark masses and infinite volume are of practical importance. An effective field theory (EFT) approach such as chiral perturbation theory (ChPT) is ideal for this purpose, as EFT only relies on the symmetries and the scale separation of the system, hence the results are model independent. 

 ChPT has been successfully applied to many aspects of meson~\cite{Gasser:1983yg}, single-~\cite{Jenkins:1990jv,Bernard:1995dp}, and
multi-nucleon systems~(see \cite{Beane:2000fx,Beane:2001bc,Bedaque:2002mn,Kubodera:2004zm,Meissner:2015una,Hammer:2012id} for reviews).
In particular, ChPT has been applied to PDF's in the meson and single nucleon systems, first in Refs.~\cite{Arndt:2001ye,Chen:2001eg,Chen:2001nb} then in~\cite{Detmold:2001jb,Detmold:2002nf,Detmold:2003tm,Detmold:2005pt,Hagler:2007xi,Gockeler:2003jfa} with more applications in PDF's
as well as other light-cone dominated observables such as generalized parton distributions (GPD's)~\cite{Chen:2001pva,Belitsky:2002jp,Chen:2003fp,Chen:2006gg,
Ando:2006sk,Diehl:2006ya}. ChPT has also been applied to multi-nucleon sectors to study the EMC effect~\cite{Chen:2004zx,Beane:2004xf} and the connection between the EMC effect and short range correlations \cite{Chen:2016bde,Lynn:2019vwp}. 

In this work, we establish the procedure to apply ChPT to LaMET. The previous success of ChPT can then be directly carried over to LaMET straight forwardly. As an example, we work out the light quark mass dependence and finite volume corrections to nucleon quasi-PDF's. Other applications such as the quenched, partially quenched, and mixed action artifacts, generalizing from SU(2) to SU(3), as well as the off-forward kinematics study of GPD's and so on, can all be studied within this framework.

\section{Applying ChPT to Quasi-PDFs}
In this section, we apply ChPT to both unpolarized and polarized isovector nucleon quasi-PDF's. The application to other quasi-observables can follow the same procedure. 

For the unpolarized nucleon quasi-PDF, the equal-time correlator computed on the lattice is 
  \begin{equation}\label{eq:unpolarized_h}
 h(z, P_z)= \frac{1}{2P_z}\langle N(P)|\bar{\psi}(z)\gamma^z W(z,0)\psi(0)|N(P)\rangle ,
  \end{equation}
where $|N(P)\rangle$ is a nucleon state with momentum $P^{\mu}=(\sqrt{M^2+P^2_z}, 0,0,P_z)$ and the 
Wilson line is
  \begin{equation}
  W(z,0)=\text{exp}\left[ig_s\lambda^\mu\int_0^zdz'A_\mu(z'\lambda^\nu)\right] ,
  \end{equation}
with $g_s$ is the strong coupling constant and $\lambda^\mu=(0,0,0,-1)$. The Fourier transform of this correlator yields the unpolarized quasi-PDF 
  \begin{equation}
\tilde{q}(x, P_z)=P_z\int_{-\infty}^{\infty}\frac{dz}{2\pi}e^{ix P_z z}\ h(z, P_z) .
  \end{equation}
 Using $\gamma^{t}$ instead of $\gamma^{z}$ in Eq.(\ref{eq:unpolarized_h})
to avoid mixing with another operator of the same mass dimension \cite{Constantinou:2017sej,Chen:2017mie} will not affect the chiral and finite volume corrections computed in this work.

 For the longitudinally polarized quasi-PDF, the equal-time correlator computed on the lattice is 
 \begin{equation}
  \label{Delta h}
  \Delta h(z,P_z)=\frac{1}{2Ms^z}\langle N(P,s)|\bar{\psi}(z)\gamma^z\gamma^5W(z,0)\psi(0)|N(P,s)\rangle ,
  \end{equation}
with the nucleon polarization vector $s^\mu=(P_z,0,0,\sqrt{M^2+P^2_z})/M$. And the corresponding quasi-PDF for quark helicity distribution is
    \begin{equation}
 \Delta\tilde{q}(x,P_z)=P_z\int_{-\infty}^{\infty}\frac{dz}{2\pi}e^{ixP_z z}\Delta h(z,P_z) .
  \end{equation}
  
  For the transversely polarized quasi-PDF, the equal-time correlator computed on the lattice is 
  \begin{equation}
  \label{delta h}
\delta h(z,P_z)=\frac{1}{2P_z s^x}\langle N(P,s)|\bar{\psi}(z)\gamma^x\gamma^z\gamma^5W(z,0)\psi(0)|N(P,s)\rangle ,
  \end{equation}
  with the nucleon polarization vector $s^\mu=(0,1,0,0)$. The corresponding quasi-PDF for quark helicity distribution is
    \begin{equation}
 \Delta\tilde{q}(x,P_z)=P_z\int_{-\infty}^{\infty}\frac{dz}{2\pi}e^{ixP_z z}\Delta h(z,P_z) .
  \end{equation}
 Replacing $\gamma^z$ in Eq.(\ref{delta h}) by $\gamma^t$ 
to avoid mixing with another operator of the same mass dimension \cite{Constantinou:2017sej,Chen:2017mie} will not affect the chiral and finite volume corrections computed in this work. 
 
Under the operator product expansion (OPE), the quark bilinear operators become
\begin{equation}
\lambda_\mu\bar{\psi}(z)\Gamma^\mu W(z,0)\psi(0)\simeq\sum_{n=0}^\infty \frac{(iz)^n}{n!}\lambda_\mu \lambda_{\mu_1}\lambda_{\mu_2}\dots \lambda_{\mu_n}\bar{\psi}\Gamma^{\mu} iD^{\mu_1}iD^{\mu_2}\dots iD^{\mu_n}\psi ,
\end{equation}
with $\lambda_{\mu}\Gamma^\mu=\gamma^z$, $\gamma^z\gamma^5$, $\gamma^x_\perp\gamma^z\gamma^5$ for the unpolarized, helicity and transversity cases, respectively. The  $\lambda_\mu \lambda_{\mu_1}\lambda_{\mu_2}\dots \lambda_{\mu_n}$ tensor is symmetric but not traceless. But the nucleon matrix elements of the trace parts are $\mathcal{O}(M^2/P_{z}^{2}, \Lambda_{\text{QCD}}^2/P_{z}^{2})$ corrections whose sizes are power suppressed \cite{Ji:2013dva,Chen:2016utp}. Therefore, we only need to concentrate on the symmetric traceless parts: 
\begin{equation}
\label{q}
  \mathcal{O}_q^{\mu\mu_1\mu_2\dots\mu_n}=\bar{\psi}\gamma^{(\mu} iD^{\mu_1}iD^{\mu_2}\dots iD^{\mu_n)}\psi ,
  \end{equation}
    \begin{equation}
    \label{Dq}
  \Delta\mathcal{O}_{q}^{\mu\mu_1\mu_2\dots\mu_n}=\bar{\psi}\gamma^{(\mu}\gamma^5 iD^{\mu_1}iD^{\mu_2}\dots iD^{\mu_n)}\psi ,
  \end{equation}
  \begin{equation}
  \label{dq}
  \delta\mathcal{O}_{q}^{x\mu\mu_1\mu_2\dots\mu_n}=\bar{\psi}\gamma^{[x}\gamma^{(\mu]}\gamma^5 iD^{\mu_1}iD^{\mu_2}\dots iD^{\mu_n)}\psi ,
  \end{equation}
where $(\dots)$ means symmetrization of the enclosed Lorentz indices with the trace parts subtracted and $[\dots]$ means the enclosed indices are antisymmetrized. 
\Blue{For instance, 
\begin{equation}
\mathcal{O}_q^{\mu\mu_1}=\bar\psi \gamma^{(\mu} D^{\mu_{1})}\psi =\frac{1}{2}\bar\psi \left(\gamma^\mu D^{\mu_{1}}+\gamma^{\mu_{1}} D^\mu\right)\psi-\frac{1}{d}g^{\mu{\mu_{1}}}\bar\psi \slashed{D}\psi ,
\end{equation}
with $d$ the spacetime dimension and
\begin{equation}
 \delta\mathcal{O}_{q}^{x\mu}=\bar{\psi}\gamma^{[x}\gamma^{\mu]}\gamma^5\psi=\frac{1}{2}\bar{\psi}(\gamma^{x}\gamma^{\mu}-\gamma^{\mu}\gamma^{x})\gamma^5\psi .
\end{equation}}

These operators are irreducible representations of the Lorentz group and are of the leading twist (twist-2). Their nucleon matrix elements give rise to moments of nucleon PDFs. 

We will use the technique developed in Refs.\cite{Arndt:2001ye,Chen:2001eg,Chen:2001nb} to match the quark level twist-2 operators to hadronic level operators using ChPT.
The Lagrangian of ChPT is given by \cite{Jenkins:1990jv,Bernard:1995dp}
  \begin{equation}
  \mathcal{L}=\frac{F_\pi^2}{4}\mathrm{tr}(\partial_\mu\Sigma \partial^\mu\Sigma^\dagger)+\eta\mathrm{tr}(\mathcal{M}\Sigma^\dagger+\mathcal{M}^\dagger\Sigma)+\overline{N}iv\cdot DN+2g_A\overline{N}S\cdot AN+\dots ,
  \end{equation}
  where the pion decay constant $F_\pi$ = 93 GeV, the pion field
  \begin{center} 
   $\Sigma=e^{\frac{i}{F_\pi}\Pi}$,\quad $\Pi= 
 \begin{pmatrix}
  \pi^0 &           \sqrt{2}\pi^+ \\
  \sqrt{2}\pi^- &  -\pi^0 
 \end{pmatrix}$ ,
 \end{center}
the quark mass matrix $\mathcal{M}=\text{diag}(m_{u},m_{d})$, and $\eta$ is the parameter connecting the quark mass and pion mass at the leading order. We will work in the isospin symmetric limit $m_{u}=m_{d}$. $N$ is the SU(2) doublet nucleon field. The nucleon velocity $v^{\mu}=P^{\mu}/M$ is the ratio of the nucleon momentum $P^{\mu}$ and the nucleon mass $M$. $D_\mu=\partial_\mu-iV_\mu$, where the pion vector current $V_\mu=\frac{i}{2}(u\partial_\mu u^\dagger+u^\dagger\partial_\mu u)$ and $u^2=\Sigma$. $g_A$=1.25 is the axial-vector coupling. The nucleon spin vector
 $S^\mu=\frac{i}{2}\sigma_{\mu\nu}v^\nu\gamma^5$. The axial vector current $A_\mu=\frac{i}{2}(u\partial_\mu u^\dagger-u^\dagger\partial_\mu u)$. The small expansion parameter $\epsilon$ in the perturbation theory is the ratio of the light to heavy scales in the problem. The light scales are the pion mass $m_{\pi}$ and the typical momentum transfer $q$, while the heavy scales are the nucleon mass $M$ and the induced scales $\Lambda_{\chi}=4 \pi F_{\pi}$ arising from the loop expansion \cite{Jenkins:1990jv}. 
 
 Although the $1/M$ expansion of ChPT looks like a non-relativistic expansion, the ChPT formulation is actually fully relativistic. The expansion requires the nucleon momentum $M v+ k$ has small off-shellness which means $2 v\cdot k +k^{2}/M \ll M$, but there is no restriction on $v$. Therefore ChPT can still be applied in our analysis where the nucleon is relativistic. 

\begin{figure}[t]
 \centering
 \begin{subfigure}{.28\textwidth}
  \centering
  \includegraphics[width=\textwidth]{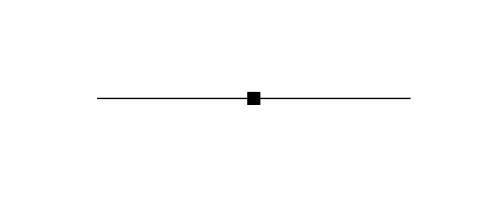}
  \caption{}
  \label{fig: chiral_1}
 \end{subfigure}
 \hfill
 \begin{subfigure}{.28\textwidth}
  \centering
  \includegraphics[width=\textwidth]{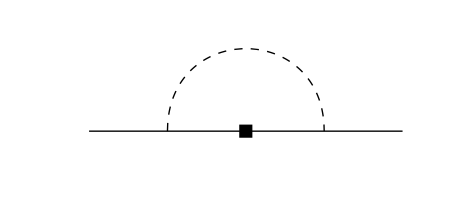}
  \label{fig: chiral_2}
  \caption{}
 \end{subfigure}
 \hfill
 \begin{subfigure}{.28\textwidth}
  \centering
  \includegraphics[width=\textwidth]{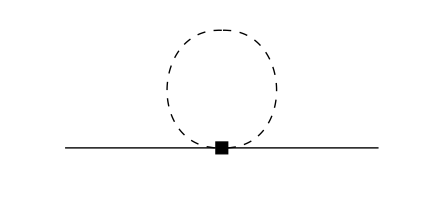}
  \label{fig: chiral_3}
  \caption{}
 \end{subfigure}
 \hfill
 \begin{subfigure}{.26\textwidth}
  \centering
  \includegraphics[width=\textwidth]{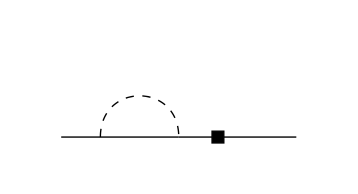}
  \label{fig: chiral_4}
  \caption{}
 \end{subfigure}
 \hfill
 \begin{subfigure}{.28\textwidth}
  \centering
  \includegraphics[width=\textwidth]{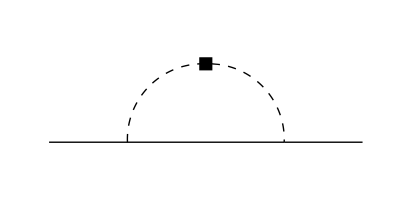}
  \label{fig: chiral_5}
  \caption{}
 \end{subfigure}
 \hfill
 \begin{subfigure}{.28\textwidth}
  \centering
  \includegraphics[width=\textwidth]{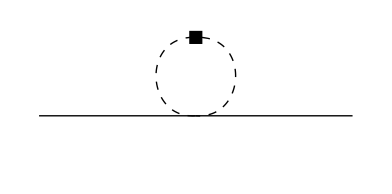}
  \label{fig: chiral_6}
  \caption{}
 \end{subfigure}
\caption{Leading Feynman diagrams in ChPT for nucleon twist-2 matrix elements. The dashed lines are pions, solid lines are nucleons, and the filled squares are the operators. Diagram (a) is the tree level diagram and diagram (d) denotes the wavefunction renormalization. Diagrams (e) and (f) only contribute to the unpolarized $n=0$ case to preserve the fermion number conservation. They are higher order in the power counting when $n>0$.}
\label{fig1}
\end{figure}

 Now, each operator in Eqs.(\ref{q})-(\ref{dq}) can be rewritten as a sum of an infinite number of hadronic operators of the same symmetries. The hadronic operators can then be organized by their mass dimensions $d_{O}$ with each operator counted as order $\epsilon^{d_{O}}$. Therefore, operators of smaller mass dimension are more important in the power counting. The isovector combinations of the lowest dimensional operators in the matching are:
  \begin{equation}
  \begin{aligned}
  \mathcal{O}_{u-d}^{\mu\mu_1\mu_2\dots\mu_n}=\ &c^{(n)}_1\bar{N}v^{(\mu} v^{\mu_1}\dots v^{\mu_n)}(u\tau^3u^\dagger+u^\dagger\tau^3u)N\\
  &+\tilde{c}^{(n)}_1\bar{N}S^{(\mu} v^{\mu_1}\dots v^{\mu_n)}(u\tau^3u^\dagger-u^\dagger\tau^3u)N + \cdots , 
  \end{aligned}
  \end{equation}
  
    \begin{equation}
  \begin{aligned}
  \Delta\mathcal{O}_{u-d}^{\mu\mu_1\mu_2\dots\mu_n}=\ &c^{(n)}_2\bar{N}S^{(\mu} v^{\mu_1}\dots v^{\mu_n)}(u\tau^3u^\dagger+u^\dagger\tau^3u)N\\
  &+\tilde{c}^{(n)}_2\bar{N}v^{(\mu} v^{\mu_1}\dots v^{\mu_n)}(u\tau^3u^\dagger-u^\dagger\tau^3u)N+ \cdots ,
  \end{aligned}
  \end{equation}
  
      \begin{equation}
  \begin{aligned}
  \delta\mathcal{O}_{u-d}^{x\mu\mu_1\mu_2\dots\mu_n}=\ &c^{(n)}_3\bar{N}S^{[x} v^{(\mu]} v^{\mu_1}\dots v^{\mu_n)}(u^\dagger\tau^3u^\dagger+u\tau^3u)N\\
  &+\tilde{c}^{(n)}_3\bar{N}S^{[x}S^{(\mu]} v^{\mu_1}\dots v^{\mu_n)}(u^\dagger\tau^3u^\dagger-u\tau^3u)N+ \cdots .
  \end{aligned}
  \end{equation}
 \Blue{The quark level operators on the left hand sides of these equations arise from OPE's. They encode the physics below the energy scale $\Lambda$, while the Wilson coefficients encode the physics above $\Lambda$. By matching these quark level operators to the most general combinations of hadronic level operators with the same symmetries, we further introduce a scale $\Lambda_{\chi}$ below $\Lambda$. The $\tilde c_i^{(n)}$ and $c_i^{(n)}$ coefficients encode the physics between $\Lambda$ and $\Lambda_{\chi}$, while the hadronic level operators encode physics below $\Lambda_{\chi}$.
 The operators with coefficients $c_i^{(n)}$($\tilde c_i^{(n)}$) have even(odd) number of pion fields. }
These nucleon operators are shown as the filled squares in diagrams (a)-(d) of Fig. \ref{fig1}. For the unpolarized case, the $n=0$ pionic operator 
    \begin{equation}
  \begin{aligned}
  \mathcal{O}_{u-d, \pi}^{\mu}\simeq a^{(0)}F_{\pi}^2\mathrm{tr}\left(\Sigma^\dagger\tau^3 i\partial^\mu \Sigma+\Sigma\tau^3 i\partial^\mu \Sigma^\dagger\right)
  \label{eq: pions}
  \end{aligned}
  \end{equation}
  can also appear in diagrams (e) and (f). They are of the same order in the power counting as diagrams (a)-(d). 
  For $n>0$, (e) and (f) become higher order diagrams.

 \section{Results}
We are interested in the finite volume effect for the nucleon quasi-PDF evaluated at nucleon momentum $P_{z}$ on a Euclidean lattice. We will work with 
a lattice with length $L$ in the three spatial directions but the size of the time direction is infinite. Assuming the nucleon and pion fields both satisfy periodic boundary conditions in the spatial directions, such that their momenta are quantized as $\vec{p}_n=2\pi\vec{n}/L$ in the reciprocal lattice space, with $\vec{n}=(n_x, n_y, n_z)$ and $n_{i}$ are integers. Poisson's formula provides a nice way to separate a discrete momentum sum into a momentum integration in the infinite volume limit and corrections caused by finite volume effect:
   \begin{equation}
  \frac{1}{L^3}\sum_{\vec{n}}f(\vec{p}_n=\frac{2\pi \vec{n}}{L})=\int\frac{\mathrm{d}^3\vec{p}}{(2\pi)^3}f(\vec{p})+\sum_{\vec{n}\neq0}\int\frac{\mathrm{d}^3\vec{p}}{(2\pi)^3}e^{i \vec{n}\cdot\vec{p}L}f(\vec{p}) .
  \end{equation}

 \begin{figure}[t]
   \centering
   \begin{subfigure}{.65\textwidth}
    \centering
    \includegraphics[width=\textwidth]{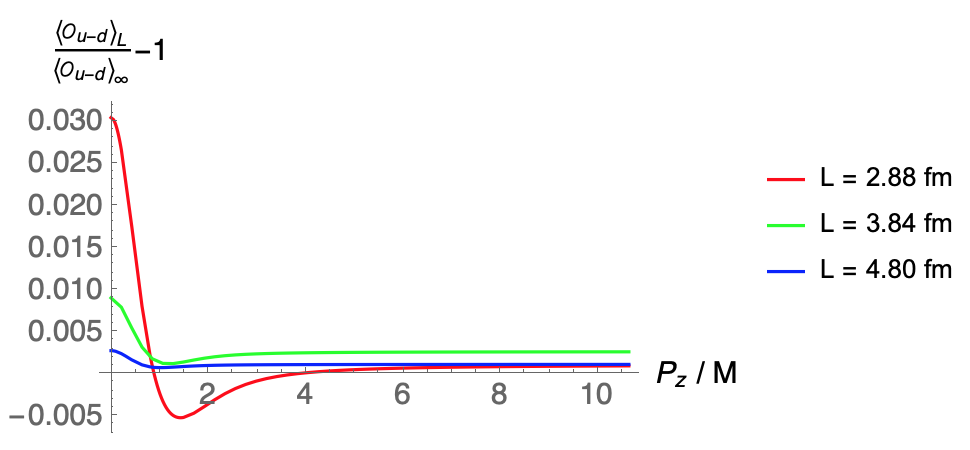}
    \caption{}
   \end{subfigure}
   \hfill
   \begin{subfigure}{.65\textwidth}
    \centering
    \includegraphics[width=\textwidth]{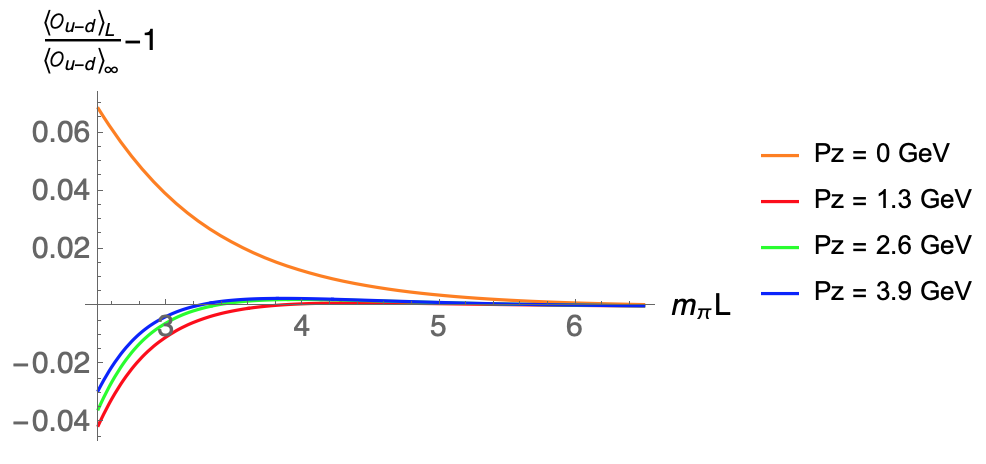}
    \caption{}
   \end{subfigure}
  \caption{Finite volume effect for the unpolarized twist-2 matrix element of Eq.(\ref{Oa}) for $m \ne 0$ shown as a function of (a) $P_{z}$ and (b) $m_{\pi} L$. The absolute value of the finite volume effect for a moving proton ($P_{z}\ne 0$) is always smaller than a rest one ($P_{z}= 0$) for any value of $m_{\pi} L$. For $P_{z}/M \ge 1$ and $m_{\pi} L \ge 3$, the finite volume effect is less than $1\%$.}
  \label{fig: finite-volume1}
 \end{figure}

  \begin{figure}[t]
   \centering
   \begin{subfigure}{.40\textwidth}
    \centering
    \includegraphics[width=\textwidth]{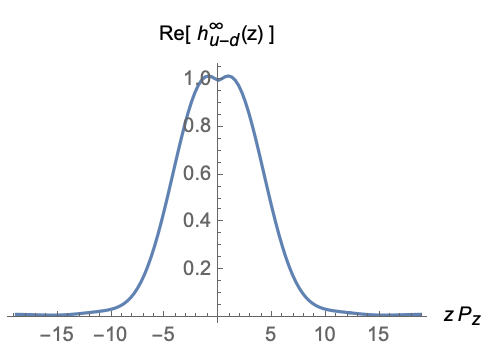}
    \caption{}
   \end{subfigure}
   \hfill
   \begin{subfigure}{.40\textwidth}
    \centering
    \includegraphics[width=\textwidth]{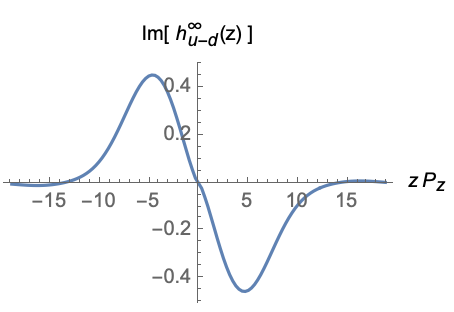}
    \caption{}
   \end{subfigure}
   \hfill
   \begin{subfigure}{.49\textwidth}
    \centering
    \includegraphics[width=\textwidth]{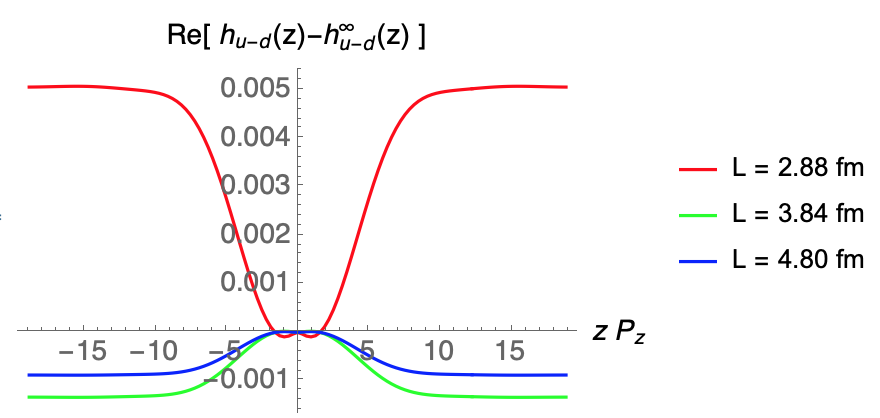}
    \caption{}
   \end{subfigure}
   \hfill
   \begin{subfigure}{.49\textwidth}
    \centering
    \includegraphics[width=\textwidth]{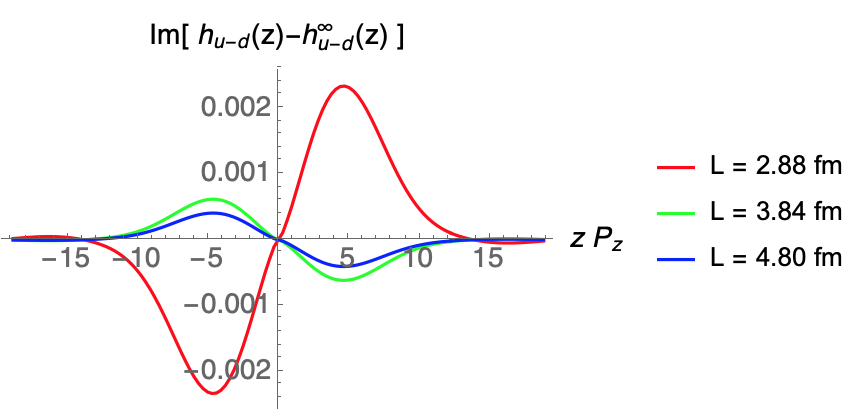}
    \caption{}
   \end{subfigure}
  \caption{The (a) real and (b) imaginary parts of the equal time correlator for an unpolarized proton in the infinite volume limit $h^{\infty}_{u-d}(z)$. This quantity is constructed using the proton isovector PDF extracted by the CTEQ-JLab collaboration (CJ12)~\cite{owens2013global} then matched to a quasi-PDF with $P_z$=1.3 GeV. The finite volume contribution is shown in (c) and (d), which is much smaller than other errors in a typical lattice QCD computation.}
  \label{fig: finite-volume}
 \end{figure}
 
   \begin{figure}[t]
   \centering
   \begin{subfigure}{.49\textwidth}
    \centering
    \includegraphics[width=\textwidth]{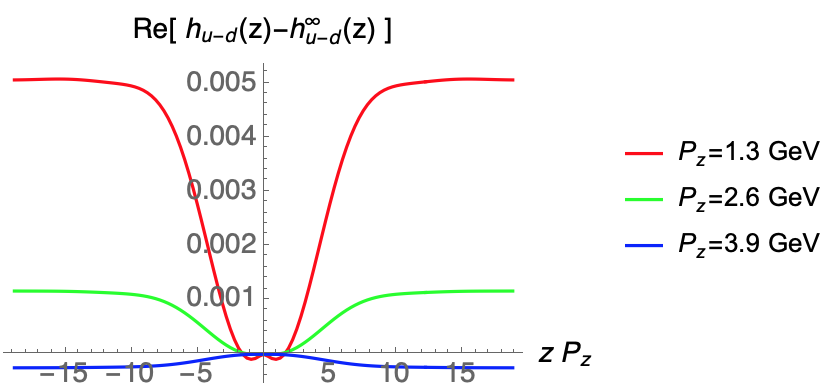}
    \caption{}
   \end{subfigure}
   \hfill
   \begin{subfigure}{.49\textwidth}
    \centering
    \includegraphics[width=\textwidth]{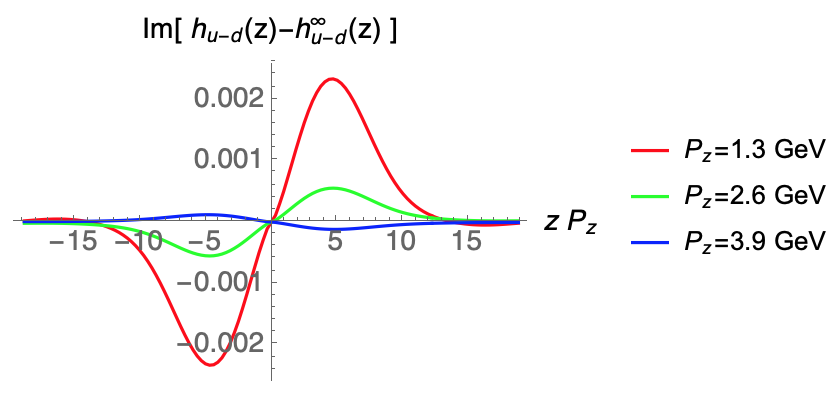}
    \caption{}
   \end{subfigure}
   \caption{Same as Fig. \ref{fig: finite-volume} but with different $P_z$'s.}
  \label{fig: finite-volumePz}
 \end{figure}
  
Our results for the nucleon twist-2 matrix elements at one loop order are
\begin{equation}
 \begin{aligned}\label{Oa}
\langle\mathcal{O}^{\mu\mu_1\cdots\mu_m}_{u-d}\rangle&=\langle\mathcal{O}^{\mu\mu_1\cdots\mu_m}_{u-d}\rangle_0\left\{1-(1-\delta_{m0})\left[\frac{3g_A^2+1}{16\pi^2F_\pi^2}m_\pi^2\log{\frac{m_\pi^2}{\Lambda_{n}^2}}+\frac{m_\pi^2}{4\pi^2F_\pi^2}\sum_{\vec{n}\neq0}\left(\frac{K_1(nm_\pi L)}{nm_\pi L}+3g_A^2 J\left(nm_{\pi}L, \frac{\vec{n}\cdot\vec{v}}{n}\right)\right)\right]\right\} , \\
\langle\Delta\mathcal{O}^{\mu\mu_1\cdots\mu_m}_{u-d}\rangle&=\langle\Delta\mathcal{O}^{\mu\mu_1\cdots\mu_m}_{u-d}\rangle_0\left[1-\frac{2g_A^2+1}{16\pi^2F_\pi^2}m_\pi^2\log{\frac{m_\pi^2}{(\Delta \Lambda_{n})^2}}-\frac{m_\pi^2}{4\pi^2F_\pi^2}\sum_{\vec{n}\neq0}\left(\frac{K_1(nm_\pi L)}{nm_\pi L}+2g_A^2 J\left(nm_{\pi}L, \frac{\vec{n}\cdot\vec{v}}{n}\right)\right)\right] , \\
\langle\delta\mathcal{O}^{\mu\mu_1\cdots\mu_m}_{u-d}\rangle&=\langle\delta\mathcal{O}^{\mu\mu_1\cdots\mu_m}_{u-d}\rangle_0\left[1-\frac{4g_A^2+1}{32\pi^2F_\pi^2}m_\pi^2\log{\frac{m_\pi^2}{(\delta\Lambda_{n})^2}}-\frac{m_\pi^2}{8\pi^2F_\pi^2}m_\pi^2\sum_{\vec{n}\neq0}\left(\frac{K_1(nm_\pi L)}{nm_\pi L}+4g_A^2 J\left(nm_{\pi}L, \frac{\vec{n}\cdot\vec{v}}{n}\right)\right)\right] , 
\end{aligned}
\end{equation}
where 
\begin{equation}
 \begin{aligned}
J\left(nm_{\pi}L, \frac{\vec{n}\cdot\vec{v}}{n}\right)&=-\frac{K_1(nm_\pi L)}{nm_\pi L}+\frac{1}{3}\left(1+\frac{(\vec{n}\cdot\vec{v})^2}{n^2}\right)K_{2}(nm_\pi L)\\
&+\int_0^\infty d\alpha \alpha(1-\cos(\alpha\vec{n}\cdot\vec{v}m_\pi L))\left[K_0(nm_\pi L\sqrt{1+\alpha^2})-\frac{1}{3}\left(1+\frac{(\vec{n}\cdot\vec{v})^2}{n^2}\right)nm_\pi L\sqrt{1+\alpha^2}K_1(nm_\pi L\sqrt{1+\alpha^2})\right] ,\\
\end{aligned}
\end{equation}
and where the subscript $0$ indicates that the corresponding matrix elements are evaluate first in the infinite volume limit, then in the chiral limit such that $m_{\pi} L \to \infty$. $n=\sqrt{n_x^2+n_y^2+n_z^2}$ and $\vec{n}\cdot\vec{v} = n_{z} P_z/M$. $n_{i}$ in Eq. (\ref{Oa}) plays the role to label the number of times the pion crosses the boundary of lattice in the $i$-direction. These matrix elements determine the $m$-th moment of the PDF defined as $\int dx x^{m}q(x)$. The $\delta_{m0}$ in the unpolarized case yields the required $(u-d)$ quark number conservation in the proton. This implies that there is a $\delta(x)$ contribution in $q(x)$ because it contributes to the zero-th moment but not any other moment.

In Eq. (\ref{Oa}), the $n$ independent part is the infinite volume result whose leading quark-mass dependence reproduces the previous results of Refs.~\cite{Chen2001, Arndt2002}. The scales $\Lambda_{n}, \Delta \Lambda_{n},  \delta \Lambda_{n}$ are associated with counterterms at the $m_{\pi}^{2}$ order that need to be fit to data. Converting from the moments to distributions in the momentum fraction $x$, both the PDF and quasi-PDF has the leading quark mass dependence
\begin{equation}
 \begin{aligned}
 \label{qa}
q_{u-d}(x)=&q_{u-d,0}(x) \left( 1-\frac{3g_A^2+1}{16\pi^2F_\pi^2}m_\pi^2\log{\frac{m_{\pi}^2}{\Lambda_{\chi}^2}} \right)+ c_{u-d,0}(x) m_{\pi}^{2} \\
&+\delta(x) \left(\frac{3g_A^2+1}{16\pi^2F_\pi^2}m_\pi^2\log{\frac{m_\pi^2}{\Lambda_{\chi}^2}} -m_{\pi}^{2} \int_{-1}^{1} d xc_{u-d,0}(x) \right) , \\
\Delta q_{u-d}(x)=&\Delta q_{u-d,0}(x) \left( 1-\frac{2g_A^2+1}{16\pi^2F_\pi^2}m_\pi^2\log{\frac{m_{\pi}^2}{\Lambda_{\chi}^2}} \right)+ \Delta c_{u-d,0}(x) m_{\pi}^{2} , \\
\delta q_{u-d}(x)=&\delta q_{u-d,0}(x) \left( 1-\frac{4g_A^2+1}{32\pi^2F_\pi^2}m_\pi^2\log{\frac{m_{\pi}^2}{\Lambda_{\chi}^2}} \right)+ \delta c_{u-d,0}(x) m_{\pi}^{2} ,
\end{aligned}
\end{equation}
where 
the functions $q, c, \Delta q, \Delta c,  \delta q, \delta c$ are $m_{\pi}$ independent. Quark number conservation $\int_{-1}^{1} d xq_{u-d}(x)=\int_{-1}^{1} d xq_{u-d,0}(x) =1$ is preserved. The delta function in $q_{u-d}(x)$ appears because we truncate the chiral expansion at one loop order. Should we go to higher loop orders, the delta function will be smeared into a more smooth function. 

In Fig.\ref{fig: finite-volume1}, we show the finite volume effect of the unpolarized twist-2 matrix elements of Eq.(\ref{Oa}) for $m \ne 0$ by taking the ratio of the matrix elements at finite and infinite volumes. We see that the finite volume effect is not monotonic in $P_{z}$ nor in $m_{\pi} L$, due to partial cancelations of several different contributions. However, the absolute value of the finite volume effect for a moving proton ($P_{z}\ne 0$) is always smaller than a rest one ($P_{z}= 0$) for any value of $m_{\pi} L$.  For $P_{z}/M \ge 1$ and $m_{\pi} L \ge 3$, the finite volume effect is less than $1\%$ and is negligible for the current precision of lattice computations. This can be interpreted as an effect due to the Lorentz contraction in the z-direction which makes the box size effectively bigger in that direction.

The finite volume effect for equal time correlators of Eqs. (\ref{eq:unpolarized_h}), (\ref{Delta h}), and (\ref{delta h}) can be derived from Eq.(\ref{Oa}): 
\begin{equation}
   \begin{aligned}\label{eq: unpolarized h}
h_{u-d}(z,P_z)&=h^{\infty}_{u-d}(z,P_z)\left[1-\frac{m_\pi^2}{4\pi^2F_\pi^2}\sum_{\vec{n}\neq0}\left(\frac{K_1(nm_\pi L)}{nm_\pi L}+3g_A^2 J\left(nm_{\pi}L, \frac{\vec{n}\cdot\vec{v}}{n}\right)\right)\right]\\
&+\frac{m_\pi^2}{4\pi^2F_\pi^2}\sum_{\vec{n}\neq0}\left(\frac{K_1(nm_\pi L)}{nm_\pi L}+3g_A^2 J\left(nm_{\pi}L, \frac{\vec{n}\cdot\vec{v}}{n}\right)\right) , \\
\Delta h_{ u- d}(z,P_z)&=\Delta h^{\infty}_{ u- d}(z,P_z)\left[1-\frac{m_\pi^2}{4\pi^2F_\pi^2}\sum_{\vec{n}\neq0}\left(\frac{K_1(nm_\pi L)}{nm_\pi L}+2g_A^2 J\left(nm_{\pi}L, \frac{\vec{n}\cdot\vec{v}}{n}\right)\right)\right] , \\
\delta h_{ u- d}(z,P_z)&=\delta h^{\infty}_{ u- d}(z,P_z)\left[1-\frac{m_\pi^2}{8\pi^2F_\pi^2}m_\pi^2\sum_{\vec{n}\neq0}\left(\frac{K_1(nm_\pi L)}{nm_\pi L}+4g_A^2 J\left(nm_{\pi}L, \frac{\vec{n}\cdot\vec{v}}{n}\right)\right)\right] .
\end{aligned}
\end{equation}
To mimic the finite volume effect on an equal time correlator computed in lattice QCD, we generate the infinite volume result by using the unpolarized isovector PDF extracted by the CTEQ-JLab collaboration (CJ12)~\cite{owens2013global}. The procedure is to perform the matching from PDF defined in the $\overline{\text{MS}}$ scheme to the quasi-PDF defined in the RI/MOM scheme, then we Fourier transform the quasi-PDF to produce the infinite volume equal time correlator $h_{u-d}^{\infty}(z,P_z)$. 

In Fig.\ref{fig: finite-volume} we use Eq.(\ref{eq: unpolarized h}) to show the finite volume effect in $h_{u-d}$. We have used $\alpha_s$=0.283, $p^z_R$=1.2 GeV, $\mu$=3.1 GeV, $\mu_R$=2.4 GeV, and $m_\pi$=0.220 GeV in the matching. In Fig.  \ref{fig: finite-volumePz}, dependence on different $P_{z}$'s is shown.  Again, these figures show that finite volume effect is negligible for current lattice computations of quasi-PDFs. We have used similar parameters as the lattice calculation of Ref.\cite{Lin2019}, where the size of finite volume effect is found to be smaller than the error of the calculation and is consistent with our result within errors. 

\Blue{Eq.(\ref{eq: unpolarized h}) shows that the combinations of correlators $\Delta h_{u-d}(z,P_z)/\Delta h^{\infty}_{u-d}(z,P_z)-1$ and $\delta h_{u-d}(z,P_z)/\delta h^{\infty}_{u-d}(z,P_z)-1$ are especially simple. They do not depend on the correlators $\Delta h_{u-d}(z,P_z)$ nor $\delta h_{u-d}(z,P_z)$. Furthermore, they do not depend on $z$. Their dependence on $m_{\pi} L$ is very similar to Fig.\ref{fig: finite-volume1}(b) and hence will not be shown again here.} 

In other versions of heavy baryon chiral perturbation, one could include $\Delta$ resonances or generalize the formalism from SU(2) to SU(3). Some of the quark mass dependence of PDF's and GPD's is already computed in these theories. However, we do not expect the finite volume effect changes a lot by adding those heavier degrees of freedom. Therefore, our conclusion on the smallness of the finite volume effects of quasi-PDF's in these theories will stay the same.

\section{Conclusion}

LaMET enables the extraction of PDFs directly on a Euclidean lattice through a factorization theorem that relates the computed quasi-PDF's to PDF's. We have applied ChPT to LaMET to further separate soft scales, such as light quark masses and lattice size, to obtain leading model independent extrapolation formulas for extrapolations to physical quark masses and infinite volume. 

We find that the finite volume effect is reduced when the nucleon carries a finite momentum. For $P_{z}/M $>1 GeV and $m_{\pi} L \ge 3$, the finite volume effect is less than $1\%$ and is negligible for the current precision of lattice computations. This can be interpreted as a Lorentz contraction of the nucleon size in the z-direction which makes the lattice size effectively larger in that direction. We also find that the quark mass dependence in the infinite volume limit computed with non-zero nucleon momentum reproduces the previous result computed at zero momentum, as expected. 

In this work, we establish the procedure to apply ChPT to LaMET. The previous success of ChPT can then be directly carried over to LaMET straight forwardly. Other applications such as the quenched, partially quenched, and mixed action artifacts, generalizing from SU(2) to SU(3), as well as the off-forward kinematics study of GPD's and so on, can all be studied within this framework.

\vspace{1em}

\section*{Acknowledgments}
This work is partly supported by the Ministry of Science and Technology, Taiwan, under Grant No. 108-2112-M-002-003-MY3 and the Kenda Foundation.

\appendix
\section{Integrals of the finite volume corrections}

In this appendix, we show how the integrals involved in the diagrams of Fig. 1 are computed. First, diagrams (c) depends on the integral
\begin{equation}
I_1=\sum_{\vec{n}\neq0}\int\frac{d^4k}{(2\pi)^4}e^{i\vec{n}\cdot\vec{k}L}\frac{1}{k^2-m^2}=\sum_{\substack{\vec{n}\neq0 \\ n_4=0}}\int\frac{d^4k_{E}}{(2\pi)^4}e^{in\cdot k_{E} L}\frac{-i}{k_{E}^2+m^2}=-\frac{im^2}{4\pi^2}\sum_{\substack{\vec{n}\neq0 \\ n_4=0}}\frac{1}{nmL}K_{1}(nmL) ,
\end{equation}
where a Wick rotation $k^0\rightarrow ik^{4}$ to Euclidean space is performed after the first equal sign with $n^\mu=(\vec{n},n_4)$ and 
$n=\sqrt{\sum_{i=1}^4n_i^2}$. We have also used the d-dimensional integral \cite{AliKhan2004,PhysRevD.69.054010},
   \begin{equation}
\sum_{n\neq0}\int\frac{d^dk_{E}}{(2\pi)^d}e^{in\cdot k_{E} L}\frac{1}{(k_{E}^2+\Delta^2)^a}=\frac{2}{(4\pi)^{d/2}\Gamma(a)}\left(\frac{nL}{2\Delta}\right)^{a-d/2}K_{a-d/2}(n\Delta L) .
  \end{equation}

Analogously, diagrams (b), (d), and (e) depend on the integral 
\begin{equation}
\begin{aligned}
I_2&=\sum_{\vec{n}\neq0}\int\frac{d^4k}{(2\pi)^4}e^{i\vec{n}\cdot \vec{k} L}\frac{(S\cdot k)^2}{(k^2-m^2)(v\cdot k)^2}\\[5pt]
&=-8im^2\sum_{\substack{\vec{n}\neq0 \\ n_4=0}}\int_0^\infty d\alpha\alpha e^{-i\alpha m n\cdot vL}\int\frac{d^4k}{(2\pi)^4}e^{in\cdot k L}\frac{(S\cdot k)^2}{[k^2+(1+\alpha^2)m^2]^3} \\[5pt]
&=4im^2\{S^\mu,S^\nu\}\sum_{\substack{\vec{n}\neq0 \\ n_4=0}}\int_0^\infty d\alpha\frac{\alpha}{L^2} e^{-i\alpha m n\cdot vL}\frac{\partial^2}{\partial n^\mu \partial n^\nu}\int\frac{d^4k}{(2\pi)^4}e^{in\cdot k L}\frac{1}{[k^2+(1+\alpha^2)m^2]^3} \\[5pt]
&=\frac{im^2}{16\pi^{2}}\left(v^\mu v^\nu+g^{\mu\nu}\right)\sum_{\substack{\vec{n}\neq0 \\ n_4=0}}\int_0^\infty d\alpha\alpha e^{-i\alpha m n\cdot vL}\frac{\partial^2}{\partial n^\mu \partial n^\nu}\left(\frac{n}{\sqrt{1+\alpha^2}mL}\right)K_{1}(n\sqrt{1+\alpha^2}m L) ,
\end{aligned}
\end{equation}
where Wick rotation is performed and all 4-vectors are defined in Euclidean space after the first equal sign and we have used the anticommutation relation in Euclidean space:
\begin{equation}
\{S^\mu,S^\nu\}=\frac{1}{2}\left(v^\mu v^\nu+g^{\mu\nu}\right)
\end{equation}

The derivative on $K_1$ yields 
\begin{equation}
\frac{\partial^2}{\partial n^\mu \partial n^\nu}\left(\frac{n}{\sqrt{1+\alpha^2}mL}\right)K_1(n\sqrt{1+\alpha^2}m L)=-g^{\mu\nu}K_0(n\sqrt{1+\alpha^2}m L)+\frac{n^\mu n^\nu}{n}\sqrt{1+\alpha^2}m L~K_1(n\sqrt{1+\alpha^2}m L) .
\end{equation}
Therefore, the final result of $I_2$ is
\begin{equation}
\begin{aligned}
I_2&=\frac{im^2}{16\pi^{2}}\sum_{\substack{\vec{n}\neq0 \\ n_4=0}}\int_0^\infty d\alpha\alpha e^{-i\alpha m n\cdot vL}\left[-3K_0(n\sqrt{1+\alpha^2}m L)+\left(1+\frac{(n\cdot v)^2}{n^2}\right)\sqrt{1+\alpha^2}nm L~K_1(n\sqrt{1+\alpha^2}m L)\right] \\[5pt]
&=\frac{im^2}{16\pi^{2}}\sum_{\substack{\vec{n}\neq0}}\int_0^\infty d\alpha\alpha \cos(\alpha m \vec{n}\cdot \vec{v}L)\left[\left(1+\frac{(\vec{n}\cdot \vec{v})^2}{n^2}\right)\sqrt{1+\alpha^2}nm L~K_1(\sqrt{1+\alpha^2}nm L)-3K_0(\sqrt{1+\alpha^2}nm L)\right] .\\[5pt]
\end{aligned}
\end{equation}
In the final step, we drop the sine function in $e^{-i\alpha m \vec{n}\cdot \vec{v}L}$ since the summation over $\vec{n}$ cancels all the terms odd in $\vec{n}$.

\bibliography{ref}
\bibliographystyle{apsrev4-1}

\end{document}